# Concept of miniature optical pressure sensor based on coupled WGMs in a dielectric microsphere

Yu.E. Geints[1,*], I.V. Minin[2], and O.V. Minin[2,**]

[1]V.E. Zuev Institute of Atmospheric Optics SB RAS, 1 Zuev square, 634021 Tomsk, Russia
[2]Tomsk State Politechnical University, Tomsk, 36 Lenin Avenue, 634050, Russia.
*ygeints@iao.ru; **prof.minin@gmail.com

**Abstract**

We present the physical concept and sample engineering design of a new miniature pressure sensor based on the whispering gallery modes (WGMs) optically excited in a dielectric microsphere placed near a flexible reflective membrane which acts as an ambient pressure sensing element. WGMs excitation is carried out by free-space coupling of optical radiation to a microsphere. The distinctive feature of proposed sensor design is double excitation of optical eigenmodes by forward and backward propagating radiation reflected from a membrane that causes WGMs interference in particle volume. The optical intensity of resulting resonant field established in the microsphere carries information about the exact position of the pressure-loaded reflecting membrane. The sensitivity of the proposed sensor strongly depends on the quality factor of the excited resonant mode, as well as geometrical and mechanical parameters of the flexible membrane. Important advantages of the proposed sensor are miniature design (linear sensor dimensions depends only on the membrane diameter) and the absence of a mechanical contact of pressure-sensitive element with WGM resonator.

**Introduction**

A classical ambient pressure sensor is essentially a transducer of the pressure applied to its sensing element into an electrical, acoustic, microwave, or optical signal measured with a corresponding device. Pressure sensors are intensively studied up to now because of their extensive applications in mechanical, electrical, and biomedical engineering [Boyd2001]. To date, various concepts of optical pressure sensors are proposed and developed, which have obvious advantages such as miniaturization, immunity to electromagnetic interference, high sensitivity and signal transmission rate [Udd2011]. Most of the optomechanical pressure sensors are based on optical fibers [Zhu2005], interferometers (such as Mach-Zehnder [Luff1998] and Michelson [Chen1999]), or MEMS structures [Bao2005]. All of them have their disadvantages and advantages, an overview of which can be found in [Lee2003, Fraden2010, Udd2011].

A distinct family comprise the optical sensors which utilize optical high-quality resonant eigenmodes commonly referred to as the Whispering Gallery Modes (WGMs) excited within lossless dielectric bodies with high degree of geometrical symmetry [Foreman2015, Zheng2018]. Among them the pressure sensors based on ring [Zhao2012, Zhang2021], disk [Ma2017], racetrack [De2007], "bottle" [Gu2017, Sumetsky2019] and spherical [Ali2019] WGM micro-resonators are reported. WGM is an optical resonance localized close to the external boundary of the resonator and usually having a high-quality factor, i.e. a narrow linewidth [Ward2011, Wang2019]. The working principle of all optical WGM sensors is usually to measure the transmittance of a special optical feeder (optical fiber, or strip-line), used to excite and readout the WGM resonator signal while scanning the excitation frequency with a tunable laser. Once a WGM is excited, within WGM bandwidth the fiber transmittance drops dramatically, which is an indicator of tuning to the resonance. Any mechanical or thermal action applied to the WGM resonator changes its characteristics and the operating eigenfrequency which then is detected by a spectrometer. By the magnitude of WGM spectral displacement one can judge the level of external mechanical load on the resonator. WGM-based optical sensors are in principle immune to electromagnetic noise and thus can have a lower level of mechanical noise compared to widely used MEMS analogues. However, the bottleneck of these circuits is the requirement of mechanical contact between the load-sensitive sensor (membrane) and the WGM resonator that limits the scope of their applications, as well as the durability of the design.

At the same time, WGM excitation in bulk microstructures can be carried out not only by evanescent electromagnetic fields using tapered fiber or prism couplers, but also by illuminating the microresonator with free-propagating optical radiation (direct free-space coupling). Despite the lower excitation efficiency [Cai2020], the free-space coupling has undeniable advantages because it does not require precise positioning of an optical coupler converting propagating optical wave into evanescent fields and the WGM resonator. Importantly, the efficiency of WGM excitation in a microparticle by direct radiation can be significantly increased using various techniques, such as side illumination by structured focused beams [Zemlyanov2000], or using the Mie scattering when placing the resonator near a reflecting substrate [Bobbert1986].

Recently, *C. Liu et al.* [Liu2000] showed by the numerical simulations that the WGMs excitation in a dielectric sphere located near a flat dielectric substrate with contrasting refractive index is accompanied by a substantial broadening and frequency shifting of the resonances, as well as WGM intensity decrease as the particle approaches the substrate. In this case, the transverse magnetic (TM) resonant modes always demonstrate a red frequency shift, while the transverse electric (TE) resonances could be shifted toward the red and blue regions of the spectrum as the optical contrast of the substrate increases. *Luk'yanchuk et al.* [Lukyanchuk2000] numerically investigated the light scattering patterns of a spherical particle located near a dielectric substrate and found a strong dependence of the scattering amplitude and phase diagram on the gap between the sphere and the substrate. Using the charges and magnet images method in [Xifre-Perez2012] an experimental proof of the photonic interaction between a dielectric silica (Si) nanoresonator and its image behind a flat gold mirror is presented. It is shown that the scattering cross section of the silica resonant cavity is enhanced when placed near the metal mirror. Similar results are reported in the experimental work by *A. Vasista et al.* on the fluorescence of the Nile blue dye by WGM excitation in a 3 µm $SiO_2$ microsphere [Vasista2018]. The sphere was placed on a gold reflecting substrate which the fluorophore was applied on. As shown, due to the contact with a metal mirror the splitting of the azimuthal resonant modes and the multiple WGM intensity enhancement took place. Interestingly, the impact of the reflecting substrate also manifests itself in nonlinear optical interactions, e.g., at the third harmonic generation in a Si nanodisk [Yao2020].

In this paper, we propose a new conceptual design of a miniature pressure sensor based on the effect of WGM excitation in a dielectric microsphere with dimensions of the order of an illuminating wavelength (mesowavelength particle). The distinctive features of the proposed sensor design are as follows: (a) a method of WGM excitation is the free-space illumination by an optical radiation tuned to the particular WGM resonance, (b) the placement of a pressure-sensitive element is contactless avoiding the mechanical impact on WGM resonator, and (c) pressure acquisition is achieved through the WGM intensity modulation (rather than its spectral shift) using the interference of WGMs excited by forward and backward propagating optical radiation upon reflection from a loaded flexible metal membrane. Specifically, due to the presence of optical reflection there is a double excitation of WGM in a spherical resonator, first by direct and then by reflected optical radiation. The optical intensity of the resulting WGM field is mediated by the interference between forward and reflected WGMs and depends on the relative position of the loaded flexible mirror. Using the finite elements method (FEM), we simulate such a sensor operation comprised of a 2 μm titanium oxide sphere and 100 nm thick gold membrane and show high sensitivity of the proposed concept to the ambient pressure.

## Working principle of a coupled WGMs (cWGM) pressure sensor

The *modus operandi* of the prototype pressure sensor is shown in Fig. 1(a) and is based on the WGMs coupling being excited in a dielectric spherical microparticle (or microcylinder) by freely propagating optical radiation when reflected from a flexible mirror (reflecting membrane). WGMs interference can be constructive or destructive depending on the phase difference of the excited eigenmodes, which will lead to a change in the amplitude of the resulting optical field of the particle. In its turn, the phase difference of direct and reflected waves depends on the deflection magnitude of the flexible mirror arising under the action of ambient pressure excess acting from outside the mirror.

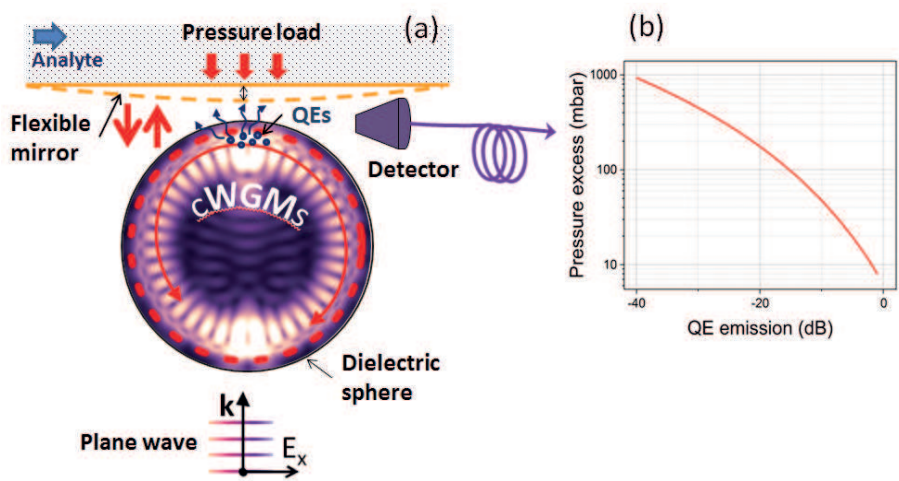

Fig. 1. (a) Schematics illustrating the physical principle of the proposed cWGM pressure sensor, (b) the model sensing working characteristic.

If some kind of quantum emitters (QEs) like quantum dots [Huber2020], spasers [Noginov2009], or nanoparticles with fluorescent substance [Lu2021, Sarkar2021, Grudinkin2015] are deposited inside the particle in the volume occupied by the resonance mode, then their emission intensity received at a photodetector will also change depending on the deflection magnitude of the sensitive mirror membrane. It is possible to choose such deflection range that the functional relation between intensity of QEs emission and mirror deflection will be unambiguous, which will allow one to reconstruct the value of measured overpressure (Fig. 1b).

## Computer simulations of cWGM excitation in a dielectric microsphere

Consider the following geometry for the excitation in a microsphere the interfering (hereinafter, coupled) "whispering gallery" resonant modes. A dielectric microsphere with the radius $R$ amounting to several optical wavelengths $\lambda$ and with refractive index $n$ is placed near a metal reflecting plate playing the role of a blind mirror and is illuminated from the back side by a plane electromagnetic wave with the amplitude $E_0$ (Fig. 2a). Note that a dielectric with high refractive index in the visible and near-infrared range can also be used as a plane mirror, such as crystalline silicon (Si) [Yue18] with RI $n \approx 4.5$ and low optical absorption at $\lambda = 1.3 \mu m$ [Aspnes83].

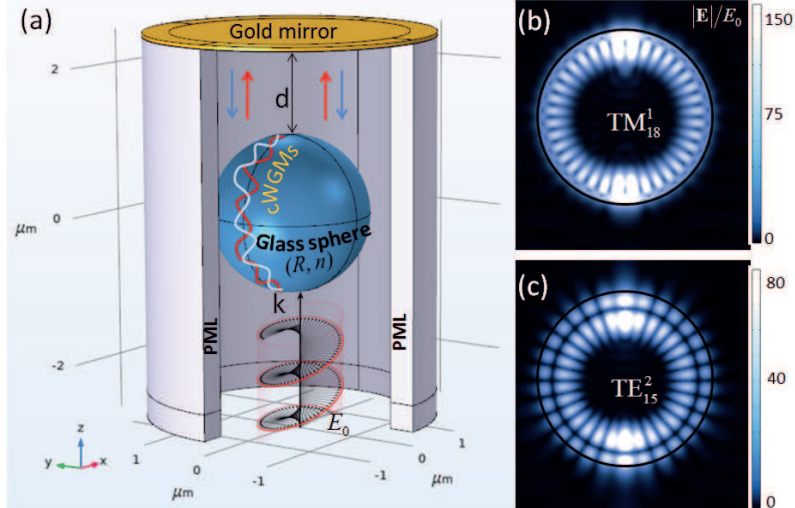

Fig. 2. (a) COMSOL model of cWGM excitation in a dielectric microsphere ($R$, $n$) placed in air and exposed to a plane circularly-polarized monochromatic optical wave (**k**) propagating at normal incidence to a flat gold mirror. Perfectly matched layers (PML) are shown surrounding the calculation domain; blue and red arrows denotes forward scattered and reflected waves, respectively; (b, c) Electric field norm |**E**| distribution in the particle cross-section attributed to different WGMs as indicated in pictures.

The electrodynamic problem of optical wave diffraction on a dielectric particle is simulated using the COMSOL Multiphysics package (ver. 5.2a) in the axisymmetric spatial geometry. The circularly polarized optical plane wave illuminating the microparticle propagates in the direction of the wave vector $\mathbf{k}_0$ along the normal to the mirror plane. Specifically, the whole photonic structure (sphere + mirror) is located in air and surrounded by a system of cylindrical perfectly absorbing layers (PML) providing the conditions of free propagation at the outer boundaries of the domain. An adaptive spatial grid of tetrahedral elements with minimum and maximum dimensions $\lambda/50$ and $\lambda/15$, respectively, is employed for the discretization of the calculation domain, providing automatic grid cells crowding in the regions with large dielectric permittivity gradient. The total number of degrees of freedom (FEM key parameter) for which the calculations is performed is of the order of $5 \cdot 10^5$.

At the optical resonance with one of the microsphere high-quality eigenmodes, a resonant configuration of the optical field is established representing many equidistantly separated intensity peaks arranged along the particle surface (Fig. 2(b,c)). Due to the presence of a plane mirror, the optical wave scattered on the particle does not leave the simulation region, but experiences a reflection which causes wave propagation back to the microparticle and re-excitation the same resonance. The optical field interference between direct and backward-reflected WGMs depending on their relative phase shift leads to the enhancement or suppression of the resulting coupled WGMs.

As an example, Figs. 2(b,c) show the electric field norm distribution of two WGMs excited in the dielectric sphere, which differ in polarization as well as azimuthal $m$ and radial indices $l$ ( $TM_m^l$, $TE_m^l$ ). A TiO$_2$ sphere with refractive index $n = 2.45$ and radius $R = 2$ µm is used to excite two

WGMs with the wavelengths $\lambda_{ml} = 1.3078562$ µm ($TM^1_{18}$) and $1.30902$ µm ($TE^2_{15}$). The field distributions are calculated without a mirror.

As noted above, a resonant internal field configuration is realized at the frequency tuning of incident light wave to the frequency of one of the sphere eigenmodes. In this case the spatial structure of the internal optical field is strongly rearranged, leading to sharp intensity enhancement and field localization close to the spherical rim with the formation of annular periodic structures in the shape of standing waves. In the framework of geometric optics, these near-surface resonance modes correspond to stable congruences of light rays refracted at a spherical surface provided that the condition of total internal reflection is fulfilled [Roll2000]. These geometrical rays are captured by the dielectric particle and propagate along its surface forming a closed spatial region bounded on one side by the inner ray caustics and on the other side by the particle surface.

Considering Figs. 2(b,c), the physical meaning of eigenmode indices becomes clear. Thus, the mode number *m* is equal to a half of the number of internal field maxima along the azimuthal angle φ, and mode order *l* corresponds to the number of maxima in radial direction *r*. Such type of WGM field distribution is predicted by the Lorenz-Mie scattering theory, which uses the internal and external electromagnetic fields representation in a spherical scatterer as the infinite series through the specific eigenmodes, so-called Mie series [Minin2015]. In the framework of the Lorenz-Mie theory the expression, e.g., of the field inside a spherical particle illuminated by a plane optical wave has the following form:

$$\mathbf{E}(\mathbf{r}) = \frac{E_0}{2kr} \sum_{m=1}^{\infty} \left\{ c_m \mathbf{M}_{1m}(\theta,\varphi) \psi_m(kr) + \frac{1}{k} d_m \nabla \times \left[ \mathbf{M}_{1m}(\theta,\varphi) \psi_m(kr) \right] \right\} + c.c. \quad (1)$$

Here, *k* is the wave number inside a particle, $\mathbf{M}_{1m}$ represents the spherical vector-harmonics defined in spherical coordinates $\mathbf{r} = \{r,\theta,\varphi\}$, $\psi_m$ denotes the spherical Riccati-Bessel functions, while $c_m$ and $d_m$ stay for partial harmonics amplitudes.

At a resonance, $\lambda = \lambda_{ml}$, only one term in the whole Mie series begins to dominate with the corresponding polarization state and mode number [Chýlek90]. This means that the major portion of light energy inside the particle is concentrated in the volume occupied by the corresponding resonant mode and is mainly distributed in the focusing regions connected along the particle surface by the annular modal structure. The spatial shape of this WGM is entirely determined by the angular profile of the corresponding vector-harmonics, i.e. $|\mathbf{E}_m(\theta)| \propto |\mathbf{M}_{1m}(\theta)|$ for TM-modes and $|\mathbf{E}_m(\theta)| \propto |\partial_\theta \mathbf{M}_{1m}(\theta)|$ for TE-modes. According to the definition, the spherical vector-harmonics are expressed through the associated Legendre polynomials $P_m(\theta)$, which are defined on the surface of a unit sphere and exhibit equidistant peaks similar to those shown in Figs. 2(b,c). Meanwhile, the WGM field localization in the radial direction is provided by the spherical Bessel function or its derivative of the corresponding azimuthal index *m*.

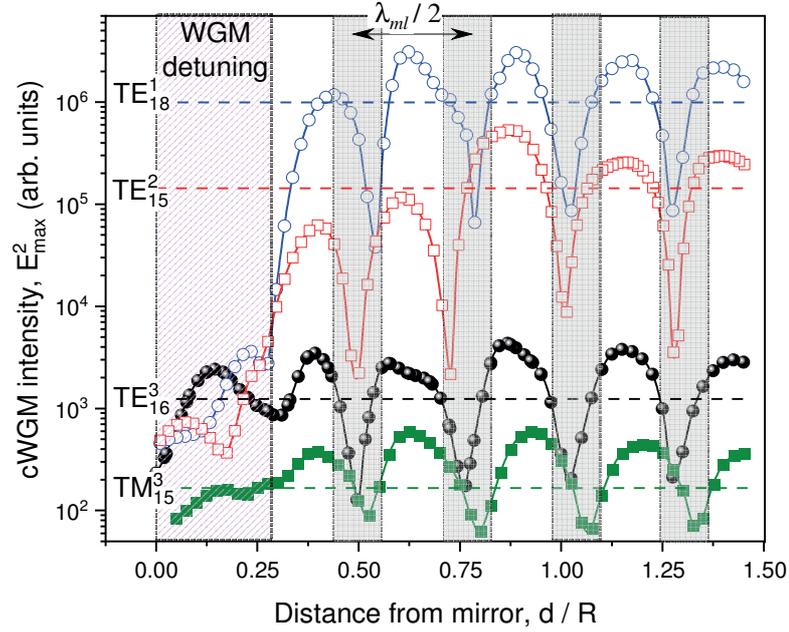

Fig. 3. cWGM peak intensity $E_{max}^2$ in TiO$_2$ sphere with $R = 2$ μm versus distance $d$ from a gold plane mirror. WGM intensity upon free-space excitation are shown by horizontal dashed lines.

Consider the excitation in a spherical microparticle the coupled WGMs provided a flat mirror is placed in the shadow particle region. In this case the interference occurs between resonant modes excited by the incident optical wave during the forward and backward propagation as shown by red and blue arrows in Fig. 2(a). For several selected modes, Fig. 3 shows the dependence of the maximal cWGM intensity $E_{max}^2$ on the normalized distance $d$ between the mirror and the particle. As seen in all cases considered, away from the mirror the intensity of the resulting resonant field regardless of its mode indices exhibits a periodic sequence of maxima and minima when the particle-mirror spacing changes. Since for each of the considered WGMs the resonant wavelength $\lambda_{ml}$ is different, the spatial coordinates of intensity extremums are also different. However, the standing wave condition is always fulfilled specifying the node-antinode alternation on the interval $\lambda_{ml}/2$.

Importantly, at the exact phase matching between coupled WGMs, the excitation of resulting eigenmode occurs with the multiple high intensity than in the classical situation of unidirectional free-space illumination of a particle by a plane wave source or a wide Gaussian beam. The values of $E_{max}^2$ in this case are shown in Fig. 3 by horizontal dashed lines for each WGM considered. Because the spatial structure of the optical wave scattered at a spherical particle and reflected back from the mirror is no longer a plane wave, the constructive interference amplification of the cWGM intensity is lowered and amounts from 2.5 to 3 times, i.e. not equal to four as in the perfect phase matched standing wave. Similarly, the destructive interference of coupled WGMs also does not lead to complete quenching of the resulting field in the nodes.

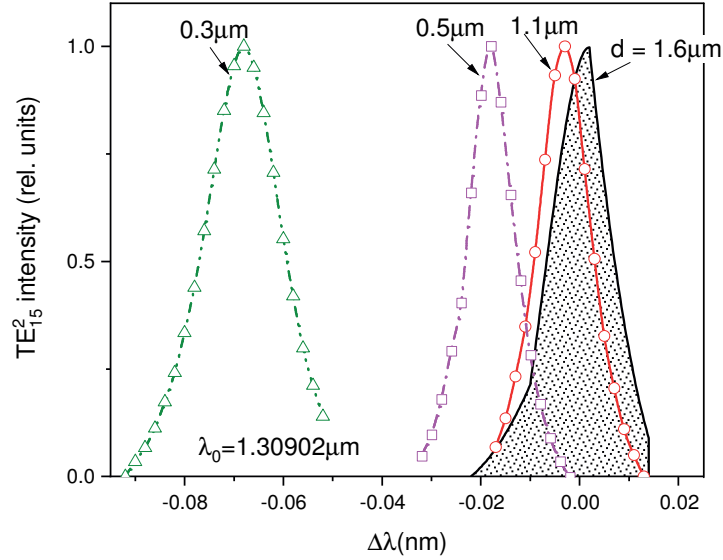

Fig. 4. Spectral shift of $TE^2_{15}$-resonance excited in a 2 μm $TiO_2$ microsphere placed at different distances *d* from a metal mirror.

Directly near the reflecting mirror, at the distances $d \approx \lambda_{ml}/2$ one can see a sharp decrease of cWGM amplitude excited in the microsphere which is especially noticeable for high-quality resonances. In this conditions, the resonant eigenmode coupling with a reflecting substrate results in a spectral shift of the resonance line from its free-space spectral position [Liu2000]. This cWGM spectral shifting is illustrated in Fig. 4 showing the spectral contours of $TE^2_{15}$ eigenmode excited in a $TiO_2$ microsphere placed at different distances from a gold plane mirror. Here, on the *x*-axis the frequency detuning $\Delta\lambda = \lambda - \lambda_0$ is shown relative to the frequency of the "unperturbed" WGM $\lambda_0$. Besides, for each particle position the optical intensity is normalized to its maximum for clarity. As follows from this figure, in the case of highly reflective metal substrate the blue shift of the resonance line and also the broadening of the resonance contour occur. Consequently, assuming the registration of light emission from a WGM is provided at the wavelength $\lambda_0$, the received intensity will gradually decrease due to the shifting of the resonance toward the short-wavelength region. Below we show how this effect can be used for the engineering a miniature ambient pressure sensor.

## Conceptual design of a cWGM ambient pressure sensor

Now consider the exemplary design of a pressure sensor exploiting the effect of coupled WGMs excitation. The sensor (see Fig. 5(a)) consists of a dielectric microsphere rigidly fixed inside a hollow cylinder having an optical window on the bottom and closed by a flexible (metal) membrane on its top possessing high reflection in the working wavelength range (optical, terahertz, etc.). The microsphere can be made of any dielectric material and doped with some light emitting nano-agent (QE), such as quantum dots, which converts the highly localized WGM field into the outgoing radiation that can be received by a photodetector via a tapered optical fiber. The cylindrical pressure sensor head can be embedded in a microchannel with an analyte which pressure is measured.

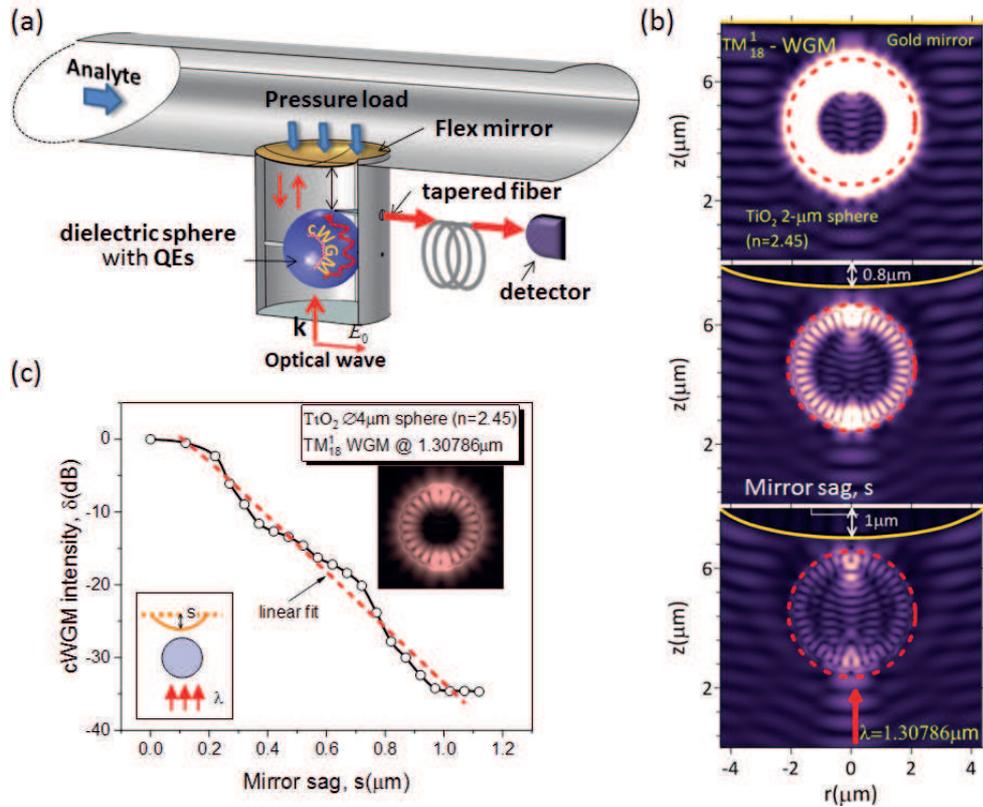

Fig. 5. (a) Principal design of the proposed cWGM-based pressure sensor; (b) cWGM intensity spatial distribution in a TiO$_2$ microsphere (2 µm) at different sags $s$ of reflecting membrane; (c) cWGM maximal intensity ($TM_{18}^1$) versus mirror sag $s$.

Through the optical window the microsphere is illuminated by optical radiation at the wavelength of one of selected WGMs. Under normal conditions, in the absence of a pressure difference inside the cylindrical sensor head and the microchannel with the analyzed substance ($\Delta p = 0$), the flexible reflecting membrane of the sensor remains flat and the received QE signal from the WGM volume is maximal. In the case of a pressure excess in the external microchannel ($\Delta p > 0$), the membrane deflects inside the sensor volume, i.e. experiences a sag, which causes the intensity decrease of coupled WGMs and a similar drop in the QE emission received by the photodetector (Fig. 5(b)). According to the dependence of cWGM intensity on the mirror sag $s$ obtained from the numerical calculations we can calculate the magnitude of the overpressure on the sensor membrane. Similar dependence is shown in Fig. 5(c) as the relative change of maximal cWGM intensity $\delta = E_{max}^2(s)/E_{max}^2(s=0)$ *versus* the deflection magnitude of the flexible mirror.

Worthwhile noting, according to our calculations, the change in cWGM intensity at the reverse deflection of the membrane toward the microchannel, which occurs at pressure deficit inside the sensor head ($\Delta p < 0$), is less sensitive to the values of $s$ due to the oscillatory dependence of $E_{max}^2$ on the distance $d$ to the mirror (see Fig. 2). This limits the use of the proposed sensor design only to the case of the ambient overpressure.

One can also propose using not cWGM intensity but resonance line spectral shift as an indicator of membrane deflection magnitude as in Fig. 4. However, the quality factor of typical WGM excited in a spherical particle is rather high and usually may exceed $10^3$ (see Fig. 2).

Consequently, WGM spectral width and its spectral shift are fractions of a nanometer that requires ultrahigh spectral resolution equipment for the registration.

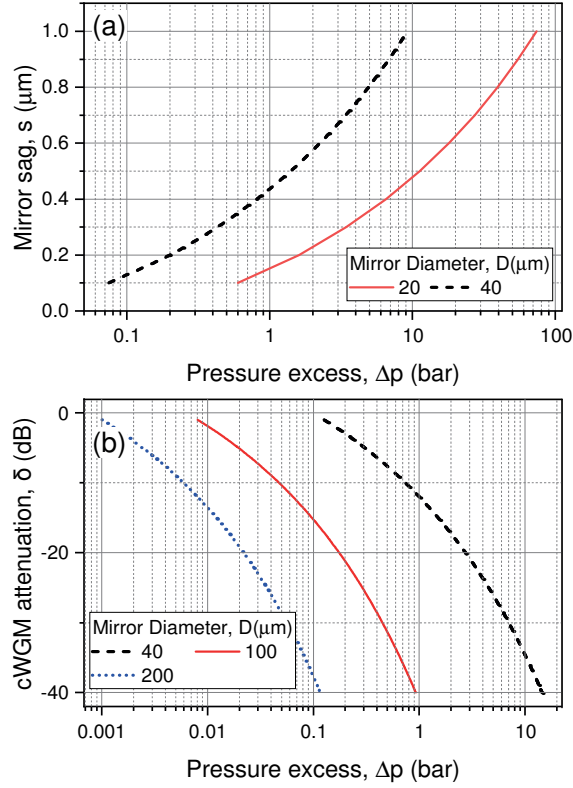

Fig. 6. Sensor operating characteristic curves utilizing $TM_{18}^{1}$ WGM and different diameter $D$ of the reflecting membrane; (a) Sensor membrane sag $s$ ($h$ = 100 nm) and (b) estimated intensity drop $\delta$ in cWGM depending on the ambient pressure excess $\Delta p$ in a microchannel.

To calculate the deflection magnitude in the center of a flat flexible membrane fixed along the contour of a cylinder under the action of an excess pressure, we use the theory of flexible plates and pillars bending [Volmir1956, Reddy200]. According to this theory, for a circular membrane with a diameter $D$ and thickness $h$, the relationship between the membrane sag $s$ and the pressure excess $\Delta p$ is calculated from the solution of the following cubic equation:

$$\frac{(\Delta p) D^4}{16 \varepsilon h^4} = \frac{16}{3(1-\mu^2)} \frac{s}{h} + \frac{6}{7}\left(\frac{s}{h}\right)^3 \quad (2)$$

Here $\varepsilon$, $\mu$ are Young's modulus and Poisson's ratio of the membrane material, respectively. The calculations of sensor membrane sag performed according to Eq. (2) under the condition $h$ = 100 nm and the membrane made of golden foil ($\varepsilon$ = 80 GPa, $\mu$ = 0.44) is shown in Fig. 6(a). Then, using the dependence of the cWGM intensity on the mirror deflection obtained from the numerical simulations (see Fig. 5(c)), one can plot the dependence of the main operating characteristic of the proposed pressure sensor. Such a characteristic curve for $TM_{18}^{1}$-mode of a dielectric sphere is shown in Fig. 6(b) for three different sensor membrane diameters $D$. As seen, by changing this parameter it is possible to manipulate the range of the registered pressure values.

It should be noted that except of the mechanical properties of the flexible membrane the sensitivity of the proposed pressure sensor depends on the type of WGM resonance excited in the mother particle. As follows from the comparison of Figs. 2(b) and (c), with the increase of mode order *l* a decrease in WGM intensity and mode field "deepening" inside the sphere volume occur. The latter effect dramatically affects the sensitivity of the coupled WGMs to distance change from the mirror. As our simulations show, if the microsphere is tuned to $TE_{15}^2$ WGM, the noticeable (> 2dB) changes in cWGM intensity arise only at the membrane sag *s* > 0.3 μm. For comparison, this threshold is *s* > 0.15 μm for the higher-order $TE_{18}^1$ mode (see Fig. 5(c)).

**Conclusion**

To conclude, we present a physical concept and a schematic engineering design of a new miniature optical pressure sensor based on the excitation of coupled WGMs in a dielectric microsphere. Unlike known analogues, the operating principle of the proposed sensor is based on recording the amplitude (intensity) changes in WGM emission rather than its spectral position shift. A pressure sensitive element of the sensor is a flexible mirror, which reflects incident optical radiation back to the particle, thus providing the interference of directly and back reflection excited resonance modes in particle volume. By recording the changes of coupled WGMs intensity one can determine the deflection of the reflecting membrane and calculate the ambient pressure excess.

Pressure sensitivity of the proposed sensor depends on the quality factor of the excited resonant modes, the radius of the flexible membrane and its elastic characteristics. Thus, when working with a high-quality WGM (quality factor ~ $10^5$) the asserted sensitivity for overpressure is from 3.2 dB/kPa for *D* = 200 μm to 25 dB/MPa for gold membrane diameter *D* = 40 μm. In the absence of any information in the literature on analogues of the proposed sensor concept, it seems difficult for us to compare the sensitivity parameters with known devices. Anyway, the sensitivity of our concept can be even increased several times if a more flexible membrane material is selected (having one order of magnitude lower Young's modulus), e.g., a commercial plastic with metallic sputtering. Besides, an important advantage of the proposed sensor is its miniature size, since the maximal sensor head dimensions are determined mainly by the diameter of the sensitive membrane. This allows the proposed concept to be used in sensor devices for microphotonics, microfluidics, "labs on a chip," etc.

**Funding**
Y G was supported by the Ministry of Science and Higher Education of the Russian Federation (V.E. Zuev Institute of Atmospheric Optics); I M and O M were partially supported by the Tomsk Polytechnic University Competitiveness Enhancement Program.